\documentclass[11pt,usletter]{article}
\usepackage{jcappub}
\usepackage{amsmath}
\usepackage{amsfonts}
\usepackage{amssymb}
\usepackage{graphicx}

\usepackage{ulem}
\usepackage{color}

\begin{document}

%
%
\title{Planck Limits on Non-canonical Generalizations of Large-field Inflation Models}
\author{Nina K. Stein and William H. Kinney} 
\emailAdd{ninastei@buffalo.edu}
\emailAdd{whkinney@buffalo.edu}
\affiliation{Dept. of Physics, University at Buffalo,
        the State University of New York, Buffalo, NY 14260-1500}
\date{\today}

\abstract{
In this paper, we consider two case examples of Dirac-Born-Infeld (DBI) generalizations of canonical large-field inflation models, characterized by a reduced sound speed, $c_{S} < 1$.  The reduced speed of sound lowers the tensor-scalar ratio, improving the fit of the models to the data, but increases the equilateral-mode non-Gaussianity, $f^\mathrm{equil.}_\mathrm{NL}$, which the latest results from the Planck satellite constrain by a new upper bound.   We examine constraints on these models in light of the most recent Planck and BICEP/Keck results, and find that they have a greatly decreased window of viability. The upper bound on $f^\mathrm{equil.}_\mathrm{NL}$ corresponds to a lower bound on the sound speed and a corresponding lower bound on the tensor-scalar ratio of $r \sim 0.01$, so that near-future Cosmic Microwave Background observations may be capable of ruling out entire classes of DBI inflation models. The result is, however, not universal: infrared-type DBI inflation models, where the speed of sound increases with time, are not subject to the bound.
}

\maketitle

%
%

\section{Introduction}
Inflationary cosmology \cite{Starobinsky:1980te,Sato:1981ds,Sato:1980yn,Kazanas:1980tx,Guth:1980zm,Linde:1981mu,Albrecht:1982wi} remains a uniquely successful phenomenological framework for understanding the origins of the universe, making quantitative predictions which current data strongly support \cite{Spergel:2006hy,Alabidi:2006qa,Seljak:2006bg,Kinney:2006qm,Martin:2006rs}. Inflation relates the evolution of the universe to one or more scalar \textit{inflaton} fields, the properties of which dictate the dynamics of the period of rapidly accelerating expansion which terminates locally in a period of reheating, followed by radiation-dominated expansion. The specific form of the potential for the inflaton field or fields is unknown, but different choices of potential result in different values for cosmological parameters, which are distinguishable by observation \cite{Dodelson:1997hr,Kinney:1998md}. Recent data, in particular the Planck measurement of Cosmic Microwave Background (CMB) anisotropy and polarization \cite{Ade:2015lrj,Ade:2015xua,Aghanim:2015xee} and the BICEP/Keck measurement of CMB polarization \cite{Ade:2015fwj} now place strong constraints on the inflationary parameter space, and falsify many previously viable  inflationary potentials, including some of the simplest and most theoretically attractive models. 

A particular class of models now known to be in conflict with CMB data is so-called ``large-field'' potentials, canonical scalar field models with field excursion during inflation $\Delta\phi \geq M_{\rm P}$, where $M_{\rm P} \equiv m_{\rm Pl} / \sqrt{8 \pi}$ is the reduced Planck mass. Such models predict a tensor scalar ratio $r \sim O(0.1)$ \cite{Lyth:1996im}, which conflicts with an upper bound from CMB data of $r < 0.066$ \cite{Kinney:2016qyl}. This is a result of the steepness of the potential, with 
\begin{equation}
r = 16 \epsilon,
\end{equation}
where $\epsilon$ is a parameter related to the equation of state during inflation, and is set by the first derivative of the potential during slow-roll inflation,
\begin{equation}
\epsilon \equiv \frac{3}{2}\left(\frac{p}{\rho} + 1\right) \simeq \frac{M_{\rm P}^2}{2} \left(\frac{V'\left(\phi\right)}{V\left(\phi\right)}\right)^2,
\end{equation}
where a prime represents a derivative with respect to the field $\phi$.
Large-field models generically have potentials steep enough that $\epsilon \sim O(0.01)$, resulting in the overproduction of tensors. An example is a quadratic potential, $V\left(\phi\right) = m^2 \phi^2$, which produces tensor/scalar ratio $r \sim 0.15$, and is therefore ruled out by Planck/BICEP/Keck \cite{Kinney:2016qyl}. (Previous work \cite{Baumann:2014cja,Palma:2014faa,Zavala:2014bda,Gobbetti:2015cya} has considered the effect of a {\it lower bound} on $r$ on the speed of sound inflationary model building. In this paper, we consider the effect of an {\it upper bound}, consistent with current constraints.)

The simplest way to reconcile single-field inflation with CMB data is to choose ``small-field'' potentials, for which $\Delta\phi \ll M_{\rm P}$ during inflation. An example is hilltop-type inflation models \cite{Kohri:2007gq,Martin:2013tda,Barenboim:2013wra,Coone:2015fha,Huang:2015cke,Vennin:2015egh,Barenboim:2016mmw}, for which $V'\left(\phi\right) \rightarrow 0$ near the inflationary fixed-point and $r \ll 0.1$, or plateau-type models such as Starobinsky inflation \cite{Starobinsky:1980te,Kehagias:2013mya}. Warm Inflation, which occurs in the presence of a thermal bath, may also reconcile large-field models with data \cite{Bastero-Gil:2016qru}. Another, less-studied alternative is to consider non-canonical generalizations of large-field inflation models \cite{Kinney:2007ag,Tzirakis:2008qy}. If we take a general non-canonical Lagrangian ${\mathcal L}\left(X,\phi\right)$, where
\begin{equation} 
X \equiv \frac{1}{2} g^{\mu\nu}\partial_\mu \phi \partial_\nu\phi
\end{equation}
is the standard canonical kinetic term, the speed of sound is no longer generically equal to the speed of light, but is given by
\begin{equation}
c_S^2 = \frac{{\mathcal L}_X}{{\mathcal L}_X + 2 X {\mathcal L}_{XX}},
\end{equation}
where a subscript indicates a derivative, ${\mathcal L}_X \equiv \delta {\mathcal L} / \delta X$. This affects inflationary observables, in particular the tensor/scalar ratio, which generalizes for non-canonical Lagrangians to
\begin{equation}
r = r = 16 \epsilon c_S^{\left(1+\epsilon\right)/\left(1-\epsilon\right)},
\end{equation}
so that, for $c_S \ll 1$, tensors are suppressed even for large $\epsilon$, and large-field models can be made consistent with CMB limits \cite{Tzirakis:2008qy} despite the steepness of the potential. 

Tensor suppression from non-canonical kinetic terms comes at a price, however. A small sound speed generates equilateral-mode non-Gaussianity in CMB fluctuations, with
\begin{equation}
f^{\rm equil.}_{\rm NL} \propto \frac{1}{c_S^2}. 
\end{equation}
The Planck CMB data place a strong {\it upper bound} on non-Gaussianity, with a $1\sigma$ constraint on the equilateral mode of \cite{Ade:2015ava}
\begin{equation}
f^{\rm equil.}_{\rm NL} = -4 \pm 43.
\end{equation}
This means that there is a limit on how much tensor suppression can be achieved before violating the bound on non-Gaussianity, since smaller $c_S$ means larger $f^{\rm equil.}_{\rm NL}$ \cite{Baumann:2014cja}. In this paper, we revisit calculations of tensor suppression from non-canonical terms in  power-law inflation \cite{Kinney:2007ag} and ``isokinetic'' generalizations of quadratic large-field inflation models \cite{Tzirakis:2008qy}, and derive new lower bounds on the tensor/scalar ratios for these cases. We find that the ultraviolet-type Dirac-Born-Infeld (DBI) models and isokinetic models are compatible with Planck bounds on both $r$ and $f^{\rm equil.}_{\rm NL}$ in both cases, but with a lower bound on the tensor/scalar ratio of $r \sim 0.01$, a figure within reach of near-future CMB experiments. Thus, next-generation CMB observations may rule out not only canonical large-field models, but classes of non-canonical generalization as well. Infrared-type DBI models are not subject to the bound. 

The paper is organized as follows: In Sec. \ref{sec:Review}, we review canonical scalar field theories for inflation as described by the useful horizon flow formalism \cite{Kinney:2002qn}, and the generalization of the flow equations to non-canonical theories \cite{Peiris:2007gz}. In Sec. \ref{sec:PowerLaw} we discuss observational limits on non-canonical generalizations of power-law inflation models, in Sec. \ref{sec:Isokinetic} we discuss the corresponding limits on isokinetic generalizations to quadratic large-field models, and in Section \ref{sec:SlowRoll} we discuss the general slow-roll case. Section \ref{sec:Conclusions} presents summary and conclusions. 

\section{Inflation from Scalar Fields}
\label{sec:Review}

In this section we review the useful horizon flow formalism for canonical and non-canonical scalar field Lagrangians \cite{Kinney:2002qn,Peiris:2007gz}.

\subsection{Canonical Scalar Fields}
We first consider canonical scalar field theories. Take a real scalar field $\phi$, with canonical action
\begin{equation}
S=\int d^4 x\sqrt{-g}\left[ \dfrac{1}{2}g^{\mu \nu}\partial_{\mu} \phi\partial_{\nu}\phi-V(\phi)\right],
\label{eq01}
\end{equation}
where $g^{\mu \nu}$ is a flat Friedmann-Robertson-Walker (FRW) metric,
\begin{equation}
ds^2 = dt^2-a^2(t)d\textbf{x}^2 .
\label{eq02}
\end{equation}
The associated equation of motion of the field is then 
\begin{equation}
\ddot{\phi}+3H\dot{\phi}+V'(\phi)=0,
\label{eq03}
\end{equation}
where an overdot represents a derivative with respect to coordinate time $t$, and the Hubble parameter $H$ is defined as 
\begin{equation}
H\equiv\dfrac{\dot{a}(t)}{a(t)},
\label{eq05}
\end{equation}
with equations of motion 
\begin{equation}
H^2 =  \left(\dfrac{\dot{a}}{a}\right)^2  = \dfrac{1}{3M_P^2}\left[\frac{1}{2}\dot{\phi}^2+V(\phi)\right],
\label{Canonical Scalar Equation of Motion 1}
\end{equation}
and
\begin{equation}
\left(\dfrac{\ddot{a}}{a}\right)  = -\dfrac{1}{3M_P^2}\left[\dot{\phi}^2-V(\phi)\right].
\label{Canonical Scalar Equation of Motion 2}
\end{equation}

Inflation is defined as a period of accelerating expansion, $\ddot{a}>0$, which occurs for the case of a slowly rolling field, $\dot{\phi}\ll V(\phi)$. If the field evolves monotonically in time, we can write the scale factor $a(\phi)$ and Hubble Parameter $H(\phi)$ as functions of the field $\phi$ rather than time. Equations  (\ref{eq03}), (\ref{Canonical Scalar Equation of Motion 1}) and (\ref{Canonical Scalar Equation of Motion 2}) can then be re-written exactly in the Hamilton-Jacobi form, 
\begin{eqnarray}
 V(\phi)&  = & 3M_P^2H^2(\phi)\left[1-\frac{2M_P^2}{3}\left(\dfrac{H'(\phi)}{H(\phi)}\right)^2\right],\cr
\dot{\phi} & = & -2M_P^2H'(\phi).
\label{phidot}
\end{eqnarray}
We then define the \textit{horizon flow} parameters as an infinite succession of  functions of $H(\phi)$ and its derivatives with respect to $\phi$ \cite{Kinney:2002qn,Liddle:1994dx},
\begin{eqnarray}
\epsilon &\equiv & 2 M^2_P \left(\dfrac{H'(\phi)}{H(\phi)}\right)^2, \cr
\eta &\equiv & 2 M^2_P \dfrac{H''(\phi)}{H(\phi)}, \cr
{}^\ell\lambda(\phi) & \equiv & \left( 2 M^2_P \right)^\ell\left(\dfrac{H'(\phi)}{H(\phi)}\right)^{\ell-1}\dfrac{1}{H(\phi)}\dfrac{d^{\ell+1}H(\phi)}{d\phi^{\ell+1}},
\label{eqa01}
\end{eqnarray}
where $\ell=2,...,\infty$ is an integer index. From these parametric definitions we obtain a set of canonical \textit{flow equations}\cite{Kinney:2002qn},
\begin{eqnarray}
\dfrac{d\epsilon}{dN} & = & 2 \epsilon (\eta-\epsilon), \cr
\dfrac{d\eta}{dN} & = & {}^2\lambda-\epsilon \eta, \cr
\dfrac{d ( {}^{\ell}\lambda ) }{dN} & = & \left[ \left( \ell-1 \right) \eta - \ell \epsilon \right] \left({}^{\ell}\lambda \right)+{}^{\ell+1}\lambda,
\label{CFE}
\end{eqnarray}
Where $N$ represents the number of e-folds of inflation, $a \propto   e^{-N}$, or 
\begin{equation}
dN \equiv -d \mathrm{ln} a = -Hdt =\dfrac{d\phi}{M_P\sqrt{2\epsilon(\phi)}},
\end{equation}
and we can write the first slow roll parameter as 
\begin{equation}
\epsilon=\frac{1}{H}\frac{dH}{dN}.
\end{equation}

Quantum fluctuations in the inflationary universe produce scalar and tensor metric perturbations with an approximately power-law form (see, \textit{e.g.,} Ref. \cite{Kinney:2009vz} for a review),
\begin{eqnarray}
P_R & = &\frac{1}{8 \pi^2}\dfrac{H^2}{M^2_P \epsilon}\bigg \vert_{k=aH} \propto k^{n_S-1},\cr
P_T & = &\frac{2}{\pi^2}\dfrac{H^2}{M^2_P}\bigg \vert_{k=aH}\propto k^{n_T}.
\label{eqa02}
\end{eqnarray}
While the scalar spectral index $n_S$ is defined as
\begin{equation}
n_S-1  \equiv  \dfrac{d\mathrm{ln}P_R}{d\mathrm{ln}k}= -4\epsilon + 2 \eta.
\label{eqa03}
\end{equation} 
The scalar fluctuation amplitude is a free parameter, and we define the tensor/scalar ratio $r$ as 
\begin{equation}
r \equiv \frac{P_T}{P_R}=16 \epsilon.
\label{eq08}
\end{equation}
We are particularly interested in $r$ and $n_S$, which are measurable parameters, and therefore provide an important mechanism for testing the validity of canonical models. 

\subsection{DBI Inflation}
The DBI scenario is intriguing due to its potential to achieve slow roll through a low speed of sound, rather than from the conventional dynamical friction due to the expansion itself \cite{Silverstein:2003hf}. This behavior, while permitting otherwise falsified models to produce tensor-scalar ratios low enough to potentially agree with the data, also introduces significant non-Gaussianity \cite{Alishahiha:2004eh,Chen:2006nt,Spalinski:2007qy}. Thus, the Planck limit on non-Gaussianity restricts these models and potentially permits us to distinguish between DBI inflation and other, similar scenarios. The DBI scenario postulates a Lagrangian for the inflaton $\phi$ of the form \cite{Silverstein:2003hf}
\begin{equation}
\mathcal{L}=-f^{-1}(\phi)\sqrt{1+f(\phi)g^{\mu \nu}\partial_{\mu}\phi\partial_{\nu}\phi}+f^{-1}(\phi)-V(\phi),
\label{eq08.1}
\end{equation}
where $V(\phi)$ is an arbitrary potential, and $f(\phi)$ is the inverse brane tension. Despite being fundamentally derived from string physics, we consider the DBI scenario as a purely phenomenological model defined by the Lagrangian (\ref{eq08.1}), where we take $f\left(\phi\right)$ to be an arbitrary free function. In this interpretation, the DBI scenario represents a special case of a larger class of inflationary models, called {\it k-inflation}, with non-canonical Lagrangians and time-dependent sound speed, first proposed by Armendariz-Picon, Damour, and Mukhanov \cite{ArmendarizPicon:1999rj}. 

If we again assume a 4-dimensional FRW metric, we can write the equation of motion for the inflaton field $\phi$ as 
\begin{equation}
\ddot{\phi}+\dfrac{3H\dot{\phi}}{\gamma^2}+\frac{V'(\phi)}{\gamma^3}+\dfrac{3f'}{2f}\dot{\phi}^2+\dfrac{f'}{f^2}\left(\dfrac{1}{\gamma^3}-1 \right)=0.
\label{eq09}
\end{equation}
The Lorentz factor $\gamma$ varies inversely with the speed of sound $c_S$, 
\begin{equation}
\gamma^2=\frac{1}{c^2_s},
\label{eq10}
\end{equation}
and, for a Lagrangian like Eq. (\ref{eq08.1}), is given by
\begin{equation}
\gamma(\phi) = \dfrac{1}{\sqrt{1-f(\phi)\dot{\phi}^2}}.
\label{eq10.1}
\end{equation}

Since the inverse sound speed $\gamma$ in general varies, the horizon flow Parameters (\ref{eqa01}) must take the $\gamma$-dependence into account. Consequently, we now refer to the original horizon flow Parameters (\ref{eqa01}) as the canonical $H-tower$, defined in terms of  $\gamma\left(\phi\right)$ and derivatives of $H\left(\phi\right)$, and introduce a second, modified sequence of parameters, the $\gamma-tower$,  defined in terms of both $H$ and derivatives of $\gamma\left(\phi\right)$ \cite{Peiris:2007gz}:
\begin{eqnarray}
\epsilon(\phi) &\equiv & \dfrac{2 M^2_P}{\gamma(\phi)} \left(\dfrac{H'(\phi)}{H(\phi)}\right)^2, \cr
\eta(\phi) &\equiv & \dfrac{2 M^2_P}{\gamma(\phi)} \dfrac{H''(\phi)}{H(\phi)},\cr
& \vdots & \cr
{}^\ell\lambda(\phi) &\equiv & \left(\dfrac{2 M^2_P}{\gamma(\phi)}\right)^\ell\left(\dfrac{H'(\phi)}{H(\phi)}\right)^{\ell-1}\dfrac{1}{H(\phi)}\dfrac{d^{\ell+1}H(\phi)}{d\phi^{\ell+1}};\cr
s(\phi) &\equiv & \dfrac{2 M^2_P}{\gamma(\phi)} \dfrac{H'(\phi)}{H(\phi)}\dfrac{\gamma '(\phi)}{\gamma(\phi)},\cr
\rho (\phi)  &\equiv & \dfrac{2 M^2_P}{\gamma(\phi)} \dfrac{\gamma ''(\phi)}{\gamma(\phi)},\cr
& \vdots & \cr
{}^\ell\alpha(\phi) &\equiv & \left(\dfrac{2 M^2_P}{\gamma(\phi)}\right)^\ell\left(\dfrac{H'(\phi)}{H(\phi)}\right)^{\ell-1}\dfrac{1}{\gamma(\phi)}\dfrac{d^{\ell+1}\gamma(\phi)}{d\phi^{\ell+1}}.
\label{eqa04}
\end{eqnarray}
We separate the H-tower and the $\gamma$-tower by a semicolon. Note that the $\gamma$-tower identically vanishes for $c_{S} = \mathrm{const.}$ The modified parameters are related by a set of first-order flow equations analogous to (\ref{CFE}) \cite{Peiris:2007gz}, 
\begin{eqnarray}
\epsilon & = & \dfrac{1}{H}\dfrac{dH}{dN}, \cr
\dfrac{d\epsilon}{dN} & = & \epsilon \left( 2\eta - 2 \epsilon -s \right) , \cr
\dfrac{d\eta}{dN} & = & - \eta (\epsilon+s) + {}^2\lambda , \cr
\dfrac{d{}^{\ell}\lambda}{dN} & = & -{}^{\ell}\lambda [\ell (s + \epsilon) -(\ell - 1)\eta]+{}^{\ell+1}\lambda , \cr
s & = & \frac{1}{\gamma} \frac{d\gamma}{dN},\cr
\dfrac{ds}{dN} & = & -s \left( 2s + \epsilon -\eta \right) +\epsilon \rho, \cr
\dfrac{d\rho}{dN} & = & - 2\rho s  + {}^2\alpha , \cr
\dfrac{d{}^{\ell}\alpha}{dN} & = & -{}^{\ell}\alpha [(\ell +1) s  +(\ell - 1)(\epsilon-\eta)]+{}^{\ell+1}\alpha. 
\label{PLMEBS02}
\end{eqnarray} 
The Hamilton-Jacobi equations for $V\left(\phi\right)$ and $f(\phi)$ generalize to
\begin{equation}
V(\phi) = 3M^2_PH^2(\phi)\left(1-\dfrac{2\epsilon(\phi)}{3}\dfrac{\gamma(\phi)}{\gamma(\phi)+1}\right)
\label{eq11}
\end{equation}
and 
\begin{equation}
f(\phi) = \dfrac{1}{2M_P^2H^2\epsilon}\left(\dfrac{\gamma^2-1}{\gamma}\right).
\end{equation}
This provides the new expression for the tensor/scalar ratio, which for $\epsilon \ll 1$ is given by
\begin{equation}
\label{eq:generalr}
r = 16 \epsilon c_S^{\left(1+\epsilon\right)/\left(1-\epsilon\right)} \simeq 16 c_S \epsilon,
\end{equation}
as well as a new form for $n_S$ \cite{Peiris:2007gz,Kinney:2007ag},
\begin{equation}
n_S-1 = -4 \epsilon +2 \eta-2s,
\label{eq13}
\end{equation}
both of which depend on the speed of sound $c_S$, the former directly and the latter through the flow parameter $s$, which we assume to be slowly varying. 

For $\epsilon$, $\eta$, and $s$ all much less than unity, equilateral-mode non-Gaussianity dominates over other modes \cite{Chen:2006nt,Cheung:2007st}, and provide a distinct observational signature of small sound speed. The Planck + BICEP/Keck constraints on $n_S$, $r$ and $f^\mathrm{equil.}_\mathrm{NL}$ are \cite{Kinney:2016qyl}:
\begin{eqnarray}
\label{eq:Planckconstraints}
r &\leq& 0.066\cr
n_S &=& 0.9644 \pm 0.0048\cr
f^\mathrm{equil.}_\mathrm{NL} &=& -4 \pm 43.
\end{eqnarray}
Since the equilateral non-Gaussianity is related to the sound speed $c_S$ by \cite{Alishahiha:2004eh}
\begin{eqnarray}
\label{eq:fNL}
f^\mathrm{equil.}_\mathrm{NL} & = & \dfrac{35}{108}\left(\dfrac{1}{c^2_S}-1\right),
\end{eqnarray}
the $2\sigma$ upper bound, $f^\mathrm{equil.}_\mathrm{NL} < 82$, on non-Gaussianity results in a $2 \sigma$ {\it lower bound} on the sound speed,
\begin{equation}
\label{eq:cSbound}
c_S \geq 0.0627.
\end{equation}
From Eq. (\ref{eq:generalr}), we see that for a given value of $\epsilon$, the lower bound on $c_S$ introduces a corresponding lower bound on the tensor/scalar ratio $r$.
We consider below several specific cases of non-canonical generalizations of large-field inflation models, and derive limits on observables.

We note that the limit (\ref{eq:cSbound}) is purely phenomenological: much tighter bounds on the sound speed can be obtained by considering UV completion of the low-energy theory. In particular, perturbative unitary \cite{Baumann:2015nta} and the presence of modified dispersion relations at high energy \cite{Baumann:2011su,Gwyn:2012mw} place much tighter bounds on the sound speed for large classes of UV-complete theories. In particular, Baumann, {\it et al.} show in Ref. \cite{Baumann:2015nta} that perturbative unitarity implies that the theory becomes strongly coupled below the scale of UV completion for $c_S > (c_S)_\star = 0.31$, a factor of five larger than the purely phenomenological limit in Eq. (\ref{eq:cSbound}). However, new weakly-coupled physics or non-perturbative effects could in principle alter this bound \cite{Baumann:2015nta}, which can only be determined with knowledge of the full UV-complete theory. For the phenomenologically derived models considered in this work, we do not in general have access to the UV-complete versions of the models, and therefore adopt the more conserative bound given in Eq. (\ref{eq:cSbound}). For the particular case of UV power-law inflation models, such as the DBI model of Silverstein and Tong \cite{Silverstein:2003hf}, considered in Sec. \ref{sec:UVPowerLaw} we also note the stronger bound derived by Baumann, {\it et al.} in our interpretation of the observational constraints. 

\section{Power-Law Models}
\label{sec:PowerLaw}
We first consider the Power-Law Inflation class of models \cite{Chimento:2007es,Kinney:2007ag}. This type of model is exactly solvable, although the current data rule out the canonical form. Constant inflationary parameters characterize this scenario,  
\begin{eqnarray}
\dfrac{d\epsilon}{dN}  = & \dfrac{d\eta}{dN} &  =  \dfrac{d^{\ell}\lambda}{dN}=0, \cr
\dfrac{ds}{dN}  = & \dfrac{d\rho}{dN} & =  \dfrac{d^{\ell}\alpha}{dN}=0.
\label{eqa06}
\end{eqnarray}
Constant $\epsilon$ implies that the Hubble parameter $H$ is exponential in $N$, 
\begin{equation}
H \propto e^{\epsilon N}.
\label{eq45}
\end{equation}
Similarly, constant $s$ implies that $\gamma$ is also exponential in $N$, 
\begin{equation}
\gamma \propto e^{sN}.
\label{eq46}
\end{equation}
By fixing $\epsilon$ and $s$, we obtain a full solution to the system of flow equations (\ref{PLMEBS02}), which provides exact solutions for the field during inflation. 

\subsection{The Canonical Case: \texorpdfstring{$\epsilon\neq0$, $s=0$, $c_S = \gamma = 1$}{Lg}}
For the canonical scenario, we take $c_S=1$ and thus $s=0$, with nonzero $\epsilon$. The solution to the canonical flow equations (\ref{CFE}) for constant parameters (\ref{eqa06}) is 
\begin{eqnarray}
\eta & = & \epsilon ,\cr
{}^{\ell}\lambda & = & \epsilon^{\ell},
\label{PLMEBS07}
\end{eqnarray}
which represents the power-law fixed point in flow space \cite{Kinney:2002qn}. Since we are working in the canonical $c_S=1$ case, we use equations (\ref{eqa03}) and (\ref{eq08}) to find $r$ and $n_S$, which are now directly related through $\epsilon$:
\begin{equation}
n_S-1=-\frac{2r}{16}=-\frac{r}{8}.
\end{equation}
Canonical power-law inflation, then, substantially overproduces tensors, with $r = 0.32$ for $n_S = 0.96$.

\subsection{Generalization}
Recalling the paired towers of inflationary flow parameters (Eqs. (\ref{eqa04})), we note the definition of $N$ as the number of e-folds before the end of inflation,
\begin{equation}
N \equiv -\int H dt =-\frac{1}{\sqrt{2M^2_P}}\int^{\phi}_{\phi_0}\sqrt{\frac{\gamma (\phi)}{\epsilon(\phi)}}d\phi,
\label{PLMEBS01}
\end{equation}
where we declare $\phi=\phi_0$ to be the end of inflation, defined as $N(\phi_0)=0$. (The constancy of $\epsilon$ in this model necessitates this alternate definition of the end of inflation, rather than the conventional $\epsilon(\phi_e)=1$.) The inflationary flow parameters (\ref{eqa04}) are related by equations (\ref{PLMEBS02}), which simplify to (\ref{CFE}) when $\gamma=1$, as the $\gamma$-tower of parameters (\ref{eqa04}) zeroes. In the general DBI case, we need both towers of parameters to completely specify the evolutionary dynamics \cite{Spalinski:2007kt}.

\subsubsection{Case 1: \texorpdfstring{$\epsilon\neq0, s=0, c_S \neq 1$.}{Lg}}
\begin{figure}
\begin{center}
\includegraphics[width=0.8\textwidth]{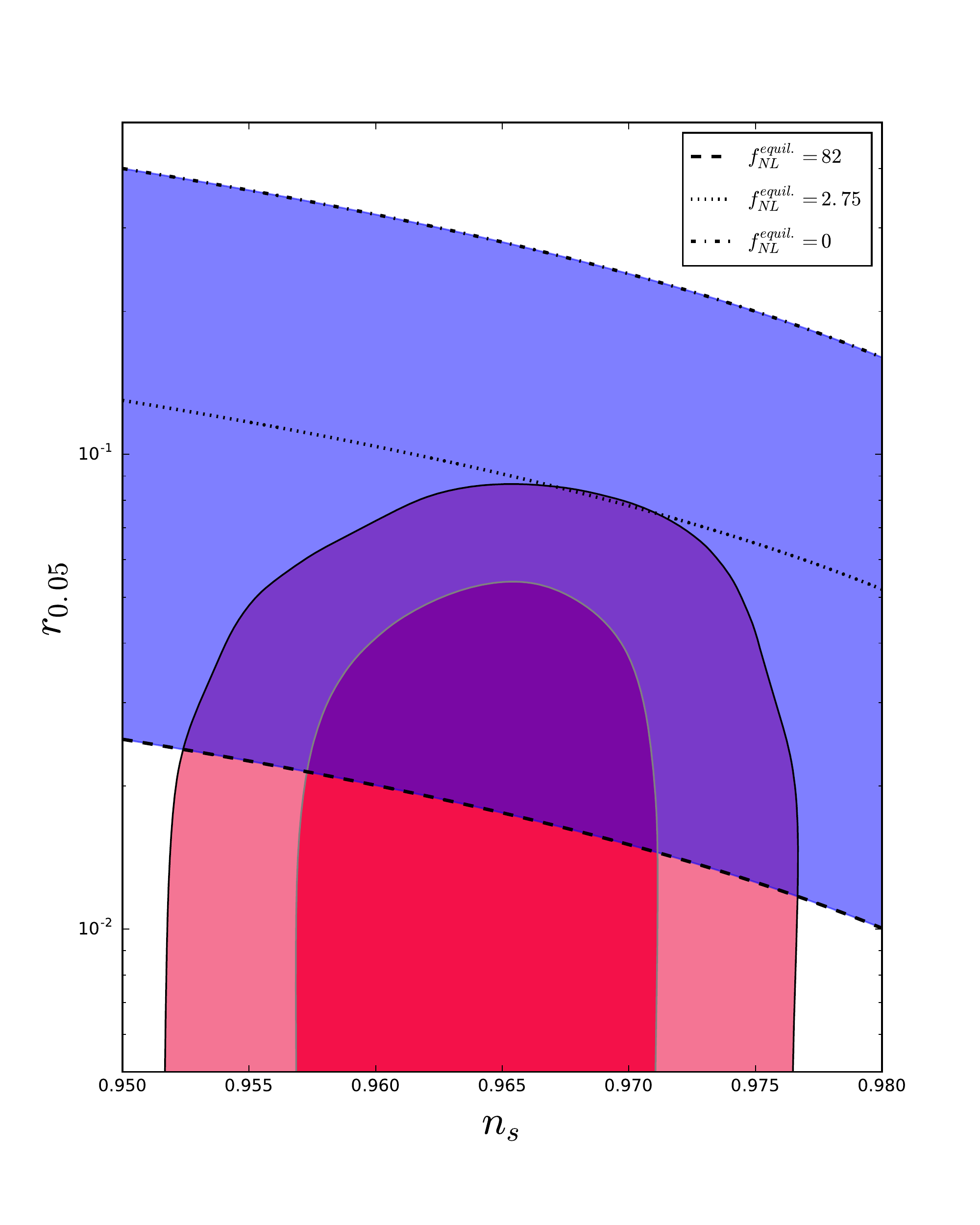}
\end{center} 
\caption{The region of viability of the Power Law Inflation scenario for the full range of potential $f_{NL}^{equil.}$ values, given constant $c_S$; the canonical case, that of $c_S=1$ and $f_{NL}^{equil.}=0$, is represented by the upper line. Here the lower bound on the sound speed results in a lower bound on the tensor/scalar ratio of $r > 0.013$ for spectral index at the upper bound $n_S = 0.9740$. We take the pivot scale for evaluating observables to be $k = 0.05\ \mathrm{h\ Mpc^{-1}}$.}
\label{Case1Plot}
\end{figure}

This is a rescaling of the canonical case, where the speed of sound remains constant while the Hubble parameter varies with time. The solution to the non-canonical flow equations (\ref{PLMEBS02}) for this scenario is identical to that of the canonical case, so that the fixed point remains (\ref{PLMEBS07}). Here, however, we use equations (\ref{eq:generalr}-\ref{eq:fNL}) for the observables, so that 
\begin{equation}
n_S-1=-\frac{r}{8c_S}.
\end{equation}
Thus, for $c_S \ll 1$, tensors are suppressed relative to the canonical version of the model. We show a comparison with Planck constraints in Fig. \ref{Case1Plot}. Additionally, as shown in \cite{Kinney:2007ag}, this case has the following solutions for $H(\phi)$, $V(\phi)$ and $f(\phi)$: 
\begin{eqnarray}
H(\phi) & = & H_0 \exp\left(-\sqrt{\dfrac{\gamma \epsilon}{2M^2_P}\phi}\right),\cr
V(\phi) & = & 3M_P^2 H_0^2 \left(1-\dfrac{2\epsilon}{3}\frac{\gamma}{(1+\gamma)}\right) \exp\left(-\sqrt{\frac{2 \gamma \epsilon}{M_P^2}}\phi \right),\cr
f(\phi) & = & \frac{1}{2 M_P^2H_0^2}\left(\frac{\gamma^2-1}{\gamma}\right)\exp\left(\sqrt{\frac{2 \gamma \epsilon}{M_P^2}}\phi \right).
\end{eqnarray}

\subsubsection{Case 2: \texorpdfstring{$\epsilon\neq0,s\neq0$.}{Lg}}
For this case, both $\epsilon$ and $s$ are nonzero and constant, resulting in a two-parameter family of solutions to the flow equations (\ref{PLMEBS02}):
\begin{eqnarray}
\eta & = & \frac{1}{2}(2\epsilon+s),\cr
{}^2\lambda & = & \frac{1}{2}(2\epsilon +s)(\epsilon + s),\cr
{}^{\ell +1}\lambda & = & {}^{\ell}\lambda\left[\epsilon+\frac{1}{2}(\ell +1)s\right],\cr
\rho & = & \frac{3s^2}{2\epsilon},\cr
{}^2\alpha & = & \frac{3s^3}{\epsilon},\cr
{}^{\ell +1}\alpha & = & \frac{1}{2}(\ell + 3)s({}^{\ell}\alpha).
\label{PLMEBS14}
\end{eqnarray}
Thus, the sound horizon, $c_SH^{-1}$, and the Hubble length, $H^{-1}$, evolve independently. We solve for these evolutionary behaviors as functions of $\phi$, beginning by isolating that of $\gamma$ \cite{Kinney:2007ag}. First, we write $s$ in terms of the constant $\epsilon$,
\begin{equation}
s \equiv \dfrac{2M^2_P}{\gamma}\left(\dfrac{H'}{H}\right)\dfrac{\gamma '}{\gamma}=\pm M_P\sqrt{2\epsilon}\dfrac{\gamma '}{\gamma^{3/2}}=\mathrm{const.},
\label{eq47}
\end{equation}
then isolate $\gamma$, 
\begin{equation}
\dfrac{d\gamma}{\gamma^{3/2}}=\pm\dfrac{s}{M_P\sqrt{2\epsilon}}d\phi,
\label{eq48}
\end{equation}
and integrate this equation to solve for $\gamma(\phi)$,
\begin{equation}
\gamma(\phi)= \left( \dfrac{8M^2_P\epsilon}{s^2}\right)\dfrac{1}{\phi^2}.
\label{eq49}
\end{equation}
Since $\gamma \propto e^{sN}$, and all other terms in the above $\gamma$ formula are constants, this implies that 
\begin{equation}
\phi^2 \propto e^{-sN}.
\label{eq50}
\end{equation}
While $\epsilon$ is strictly positive, $s$ is not, and its sign depends on the direction of the field evolution: 
\begin{equation}
\frac{d\phi}{\phi}=-\frac{s}{2}dN,
\label{eq51}
\end{equation}
which we can rewrite as 
\begin{equation}
\frac{d\phi}{dN}=-\frac{s}{2}\phi,
\end{equation}
where we implicitly assume positive $\phi$. This means that a positive $s$-value corresponds to $\phi$ increasing as $N$ decreases, evolving from lower to higher $\phi$ and $c_S$ values, referred to as the {\it infrared (IR)} case. Conversely, a negative $s$-value represents the opposite case, where $\phi$ decreases with $N$, so that $\phi$ and $c_s$ evolve from higher to lower values, referred to as the {\it ultraviolet (UV)} case. 

We now substitute our formula for $\gamma$ into the definition of $\epsilon$ from Eqs. (\ref{eqa04}), 
\begin{equation}
\epsilon \equiv  \dfrac{2 M^2_P}{\gamma(\phi)} \left(\dfrac{H'}{H}\right)^2 = \dfrac{s^2\phi^2}{4\epsilon}\left(\dfrac{H'}{H}\right)^2=\mathrm{const.},
\label{eq52}
\end{equation}
and isolate $H$,
\begin{equation}
\dfrac{dH}{H}=\pm \dfrac{2\epsilon}{s}\dfrac{d\phi}{\phi},
\label{eq53}
\end{equation}
so that $H(\phi)$ becomes
\begin{equation}
H \propto \phi^{\pm2\epsilon/s},
\label{eq54}
\end{equation}
which reduces to 
\begin{equation}
H \propto \phi^{-2\epsilon/s} \propto e^{\epsilon N},
\label{eq55}
\end{equation}
when we take ($dH/dN > 0$), consistent with the Null Energy Condition, $p \geq -\rho$. Recalling that we are defining the end of inflation as occurring at some point $\phi_0>0$, defined as $N(\phi_0)=0$, we can define the exact solution for $\phi^2$,
\begin{equation}
\phi^2=\phi_0^2\epsilon^{-sN}.
\label{PLMEBS15}
\end{equation}
This corresponds to $V(\phi)$ and $f(\phi)$ of 
\begin{equation}
V(\phi)=3M_P^2H^2(\phi)\left(1-\dfrac{2\epsilon}{3} \dfrac{1}{1+c_S(\phi)}\right),
\label{case2potential}
\end{equation}
and 
\begin{equation}
f(\phi)=\left(\frac{1}{2M_P^2 \epsilon}\right)\frac{1-c_S^2(\phi)}{H^2(\phi)c_S(\phi)}.
\end{equation}
Note, here, that while the potential $V(\phi)$ is well-behaved for superluminal values of $c_S$, the function $f(\phi)$ becomes negative. Strictly from the standpoint of the DBI action, this is not physically inconsistent, but embedding in string theory requires $f\left(\phi\right)$ to be positive-definite \cite{Bessada:2009ns}, and we adopt a causality constraint for the purpose of this analysis. In particular, in the IR case, with $\gamma \propto e^{-s N}$, with $s > 0$, the speed of sound becomes superluminal at late times, and we set the end of inflation ($N = 0$) at the point where $c_S = 1$, so that
\begin{equation}
\gamma = e^{-s N}.
\end{equation}
In stringy language, this corresponds to the point at which the brane responsible for inflation leaves the warped throat and enters the bulk \cite{Kinney:2007ag}.
Similarly, an exact analytical solution for perturbations can be derived (as shown in \cite{Kinney:2007ag}), showing that $P(k)$ is an exact power law in $k$ and providing an expression for $n_S$,
\begin{equation}
n_S = 1-\dfrac{2\epsilon+s}{1-\epsilon-s}.
\label{PLMEBS17}
\end{equation}

\subsection{Constraints}
\begin{figure}
\includegraphics[width=\linewidth]{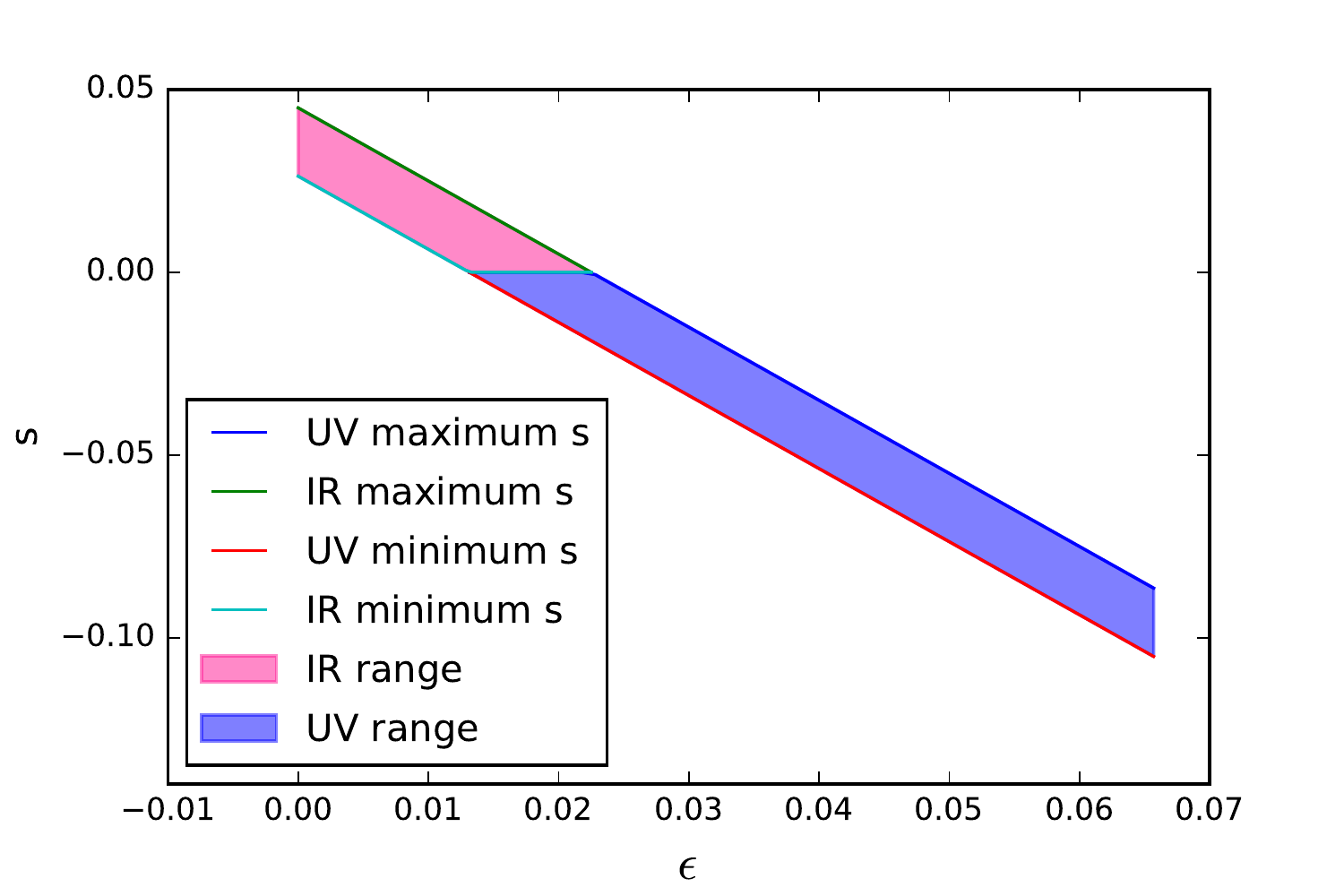} 
\caption{The acceptable range of $s$ and $\epsilon$ value combinations that lie within the $2\sigma$ range of $n_S$.}
\label{fig2}
\end{figure}

We have thus found a set of constraint equations for power-law models: 
\begin{eqnarray}
n_S -1 & = & -\dfrac{2\epsilon+s}{1-\epsilon-s},\cr
r & = & 16 c_S \epsilon, \cr
f^\mathrm{equil.}_\mathrm{NL} & = & \dfrac{35}{108}\left(\dfrac{1}{c^2_S}-1\right) .
\end{eqnarray}
We combine these with the bounds given in Eq. (\ref{eq:Planckconstraints}) to determine the viability of the power law scenarios. We combine the upper bound on $r$ with the lower bound (\ref{eq:cSbound}) on $c_S$ to find an upper bound on $\epsilon$,
\begin{equation}
\epsilon =\dfrac{r}{16c_S} \leq 0.0657.
\label{eq57}
\end{equation}
Now we use $n_S$, which current data bounds from both above and below, to find a pair of inequalities relating $\epsilon$ and $s$, 
\begin{equation}
-0.045 \leq  -\dfrac{2\epsilon + s}{1-\epsilon - s} \leq -0.0263,
\label{eq58}
\end{equation}
so that the value of $n_S$  defines the range of $s$- and $\epsilon$-values, as shown in Figure \ref{fig2}. 
For small $s$ and $\epsilon$, we simplify equation (\ref{eq58}) by approximating $1-\epsilon-s\approx 1$, 
\begin{equation}
-0.045 \leq  -2\epsilon - s \leq -0.0263, 
\label{eq59}
\end{equation}
which we rewrite as
\begin{equation}
2\epsilon - 0.045+ \leq -s \leq 2\epsilon - 0.0263
\label{isolates}
\end{equation}
The UV and IR cases result in separate limits on observables.

\subsubsection{The UV case, \texorpdfstring{$s<0$}{Lg}}
\label{sec:UVPowerLaw}
\begin{figure}
\includegraphics[width=\linewidth]{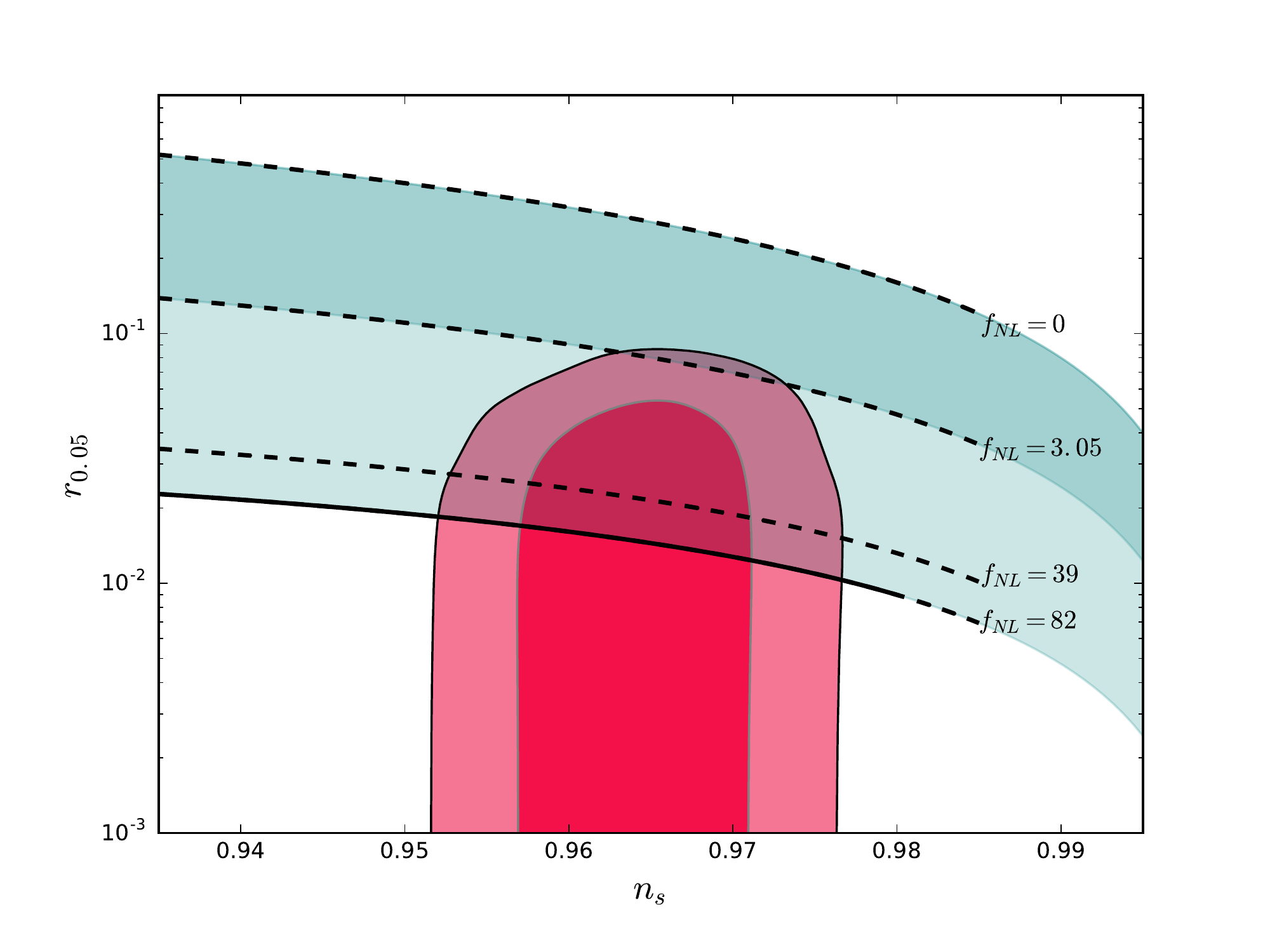} 
\caption{The range of UV DBI models for the full spectrum of $f^\mathrm{equil.}_\mathrm{NL}$ values, from 0 to 82, shown shaded in green. Isocontours of $f^\mathrm{equil.}_\mathrm{NL}$ are indicated by the dotted lines, and the pink shaded region represents the most recent Planck results. The darker shaded region represents the range allowed by perturbative unitarity, $f_{NL} < 3.05$, which is inconsistent with the region allowed by Planck.}
\label{fig3}
\end{figure}

Taking the absolute value of s in equation (\ref{isolates}), we have 
\begin{equation}
2\epsilon-0.045 \leq \left\vert s\right\vert \leq 2\epsilon-0.0263.
\end{equation} 
Since $\left\vert s \right\vert \geq 0$, the second inequality gives a lower bound on $\epsilon$, 
\begin{equation}
\epsilon>0.01315.
\end{equation}
This, combined with the upper bound $r < 0.066$ from Planck+BICEP/Keck, results in an upper bound on $c_S$,
\begin{equation}
c_S = \frac{r}{16\epsilon} \leq 0.31.
\label{eq64}
\end{equation}
This upper bound corresponds to a lower bound on $f^\mathrm{equil.}_\mathrm{NL}$ of about 2.84. In addition, the lower bounds on $\epsilon$ and $c_S$ also give a lower bound on $r$,
\begin{equation}
r=16c_s\epsilon\geq 0.0132\approx0.01,
\end{equation}
so that any improved measurements on $r$ reducing it below $\approx 0.01$ will disallow all power-law UV DBI models. Any measurements that reduce $f^\mathrm{equil.}_\mathrm{NL}$ below the 2.84 limit would also falsify these models, though we are unlikely to observationally reduce $f^\mathrm{equil.}_\mathrm{NL}$ below 2.84 before we reduce $r$ below 0.01. The region covered by this class of model in $r-n_S$ space is shown in Figure \ref{fig3}.
Here we note that the upper bound (\ref{eq64}) $c_S \leq 0.31$ is exactly the same as the lower bound $c_S > (c_S)_* = 0.31$ from perturbative unitarity derived in Ref. \cite{Baumann:2015nta}, which would then rule out this class of models in its entirety due to the presence of strong coupling below the symmetry breaking scale. This is the case for string-theory derived models such as the the original DBI model of Silverstein and Tong \cite{Silverstein:2003hf}. For this model in particular, $s = - 2 \epsilon$, which is already ruled out by the spectral index constraint (\ref{eq59}).

\subsubsection{The IR case, \texorpdfstring{$s>0$}{Lg}}
\label{sec:IRDBI}
In this case, s is positive, so taking the absolute value of $s$ inverts the  inequalities from Eq. (\ref{isolates}), 
\begin{equation}
0.045-2\epsilon \geq \left\vert s\right\vert \geq 0.0263 -2\epsilon,
\end{equation}
For $\left\vert s \right\vert \geq 0$, the first inequality gives an {\it upper} bound on $\epsilon$,
\begin{equation}
\epsilon\leq 0.0225,
\end{equation}
which means that the IR case, unlike the UV, does not have a minimum tensor-scalar ratio, and we cannot restrict it in the same way. We discuss this issue further when we consider the general slow-roll case in Sec. \ref{sec:SlowRoll}. Note that in this case, it is trivial to satisfy the condition for existence of a Bunch-Davies vacuum \cite{Kinney:2007ag},
\begin{equation}
\epsilon < 1 - s.
\end{equation}
Causality, likewise, provides no additional constraint.

\section{Isokinetic Inflation}
\label{sec:Isokinetic}

The second class of models we consider here is the isokinetic case, characterized by constant field velocity $\dot{\phi}$. This model, which is a generalization of the $V \propto \phi^2$ canonical case, is not exactly solvable, but instead requires the slow roll approximation.
\subsection{Canonical Case}
We first take a canonical Lagrangian with a quadratic potential $V(\phi)=m^2\phi^2$. Since the slow-roll limit implies that $H\propto\sqrt{V}$, it follows that $H'\left(\phi\right) = \mathrm{const.}$, and $H''=0$.  From equation (\ref{phidot}), we have $\dot{\phi}\propto H'$, which results in $\dot{\phi}=\mathrm{const.}$, and $\eta \propto H''\left(\phi\right) = 0$. All higher order flow parameters also vanish. This leaves only one nonzero flow parameter, $\epsilon$. 
As this is the canonical case, we use equations (\ref{eqa03}) and (\ref{eq08}) to find $r$ and $n_S$, related through $\epsilon$;
\begin{equation}
n_S-1=-4\epsilon=-\frac{r}{4},
\end{equation}
which again overproduces tensors inconsistent with CMB constraints.

\subsection{Non-Canonical DBI Generalization}
The non-canonical generalization of this case is more complex; the addition of a non-zero $s$ term means that the constant $\dot\phi$ solution satisfies \cite{Tzirakis:2008qy}
\begin{equation}
\ddot{\phi}=-\dot{\phi}H(\eta-s)=0,
\label{eq22}
\end{equation}
so that constant $\dot{\phi}$ requires $\eta(\phi)=s(\phi)$. This result is equivalent to requiring 
\begin{equation}
\dfrac{H''(\phi)}{H'(\phi)}= \dfrac{\gamma ' (\phi)}{\gamma (\phi)},
\label{eq23}
\end{equation}
and we can similarly prove that \cite{Tzirakis:2008qy}
\begin{equation}
\dfrac{H'''(\phi)}{H'(\phi)}= \dfrac{\gamma '' (\phi)}{\gamma (\phi)}.
\label{eq24}
\end{equation}
This pattern persists, relating the $\ell$-th derivative of $\gamma$ to the $\ell+1$-th derivative of $H$, so that the parameters of the $H$-tower relate to those of the $\gamma$-tower by
\begin{equation}
{}^{\ell+1}\lambda(\phi)={}^\ell\alpha(\phi)\epsilon(\phi),\ell=1,2,...
\label{eq25}
\end{equation}
With the flow parameter $\epsilon$ bridging the two towers. 

By allowing nonconstant sound speed, however, we raise a problem avoided in the canonical case: variable sound speed precludes an unique solution for $H(\phi)$ \cite{Tzirakis:2008qy}. Consequently, isokinetic solutions correspond to a whole \textit{class} of potentials, where
\begin{equation}
\eta(\phi)  =  s(\phi)  =\dfrac{2M^2_P B}{H(\phi)}\left(\dfrac{\gamma ' (\phi)}{\gamma (\phi)}\right),
\end{equation}
and
\begin{equation}
\epsilon(\phi) \equiv   \dfrac{2 M^2_P}{\gamma(\phi)} \left(\dfrac{H'(\phi)}{H(\phi)}\right)^2  = 2M^2_PB \dfrac{H'(\phi)}{H^2(\phi)},
\label{eq29}
\end{equation}
where $B \equiv - \dot\phi / \left(2 M_\mathrm{P}^2\right)$ is a constant.

We must then proceed by ansatz; for large $\phi$, we assume that the dominant term of $H(\phi)$ goes as $\phi^p$, so that $H(\phi)$ has the general form of 
\begin{equation}
H(\phi) = C \mu^{(1-p)}\phi^p,
\label{eq30}
\end{equation} 
where $\mu$ has dimension of mass, and C is defined as an arbitrary scaling constant $\alpha_0$ multiplied by B. The slow roll case, where $V \propto H^2 \propto \phi^{2p}$, results in a $\gamma$ function of 
\begin{equation}
\gamma (\phi) = \alpha_0 p \mu^{(1-p)}\phi^{(p-1)},
\label{eq31}
\end{equation}
and an $\epsilon$ of 
\begin{equation}
\epsilon (\phi) = \dfrac{2pM^2_P}{\alpha_0 \mu^{(1-p)}}\phi^{-(p+1)}=\left(\dfrac{\phi_e}{\phi}\right)^{p+1},
\label{eq32}
\end{equation}
where $\phi_e$ represents the value of $\phi$ at the end of inflation, as defined by $\epsilon(\phi_e) \equiv 1$, so that 
\begin{equation}
\phi_e=\left(\dfrac{2pM^2_P}{\alpha_0 \mu^{(1-p)}}\right)^{1/(p+1)},
\label{eq33}
\end{equation}
and we can rewrite the equation for $\gamma$ as 
\begin{equation}
\gamma(\phi) = 2 p^2 \left(\dfrac{M_P}{\phi_e}\right)^2\left(\dfrac{\phi}{\phi_e}\right)^{p-1}.
\label{eq34}
\end{equation}

The number of e-folds, $N(\phi)$, is given by
\begin{equation}
N(\phi)=\dfrac{1}{\sqrt{2M^2_P}} \int_{\phi_e}^{\phi} \sqrt{\dfrac{\gamma(\phi}{\epsilon(\phi)}}d\phi=\dfrac{p}{\phi_e} \int_{\phi_e}^{\phi}\left(\dfrac{\phi}{\phi_e} \right) ^p d\phi=\dfrac{p}{p+1}\left(\frac{1}{\epsilon}-1\right),
\label{eq35}
\end{equation}
allowing us to write $\epsilon$ as a function of $N$, rather than $\phi$, 
\begin{equation}
\epsilon(N)=\dfrac{1}{1+(p+1)N/p}.
\label{eq36}
\end{equation}
Using these values of $\epsilon$ and $\gamma$, we derive a formula for the tensor/scalar ratio $r$, 
\begin{equation}
r = \frac{16 \epsilon}{\gamma}=\frac{8}{p^2}\left(\dfrac{\phi_e}{M_P}\right)^2 \epsilon^{2p/(p+1)}=\frac{8}{p^2}\left(\dfrac{\phi_e}{M_P}\right)^2 \left(1+(p+1)N/p\right)^{-2p/(p+1)}.
\label{eq38}
\end{equation}
Furthermore, since $\eta=s$, the scalar spectral index $n_S$ depends on $\epsilon$ alone:
\begin{equation}
n_S-1 = -4\epsilon+2\eta-2s=-4\epsilon=\dfrac{-4}{1+(p+1)N/p}.
\label{eq37}
\end{equation}

\begin{figure}
\includegraphics[width=\linewidth]{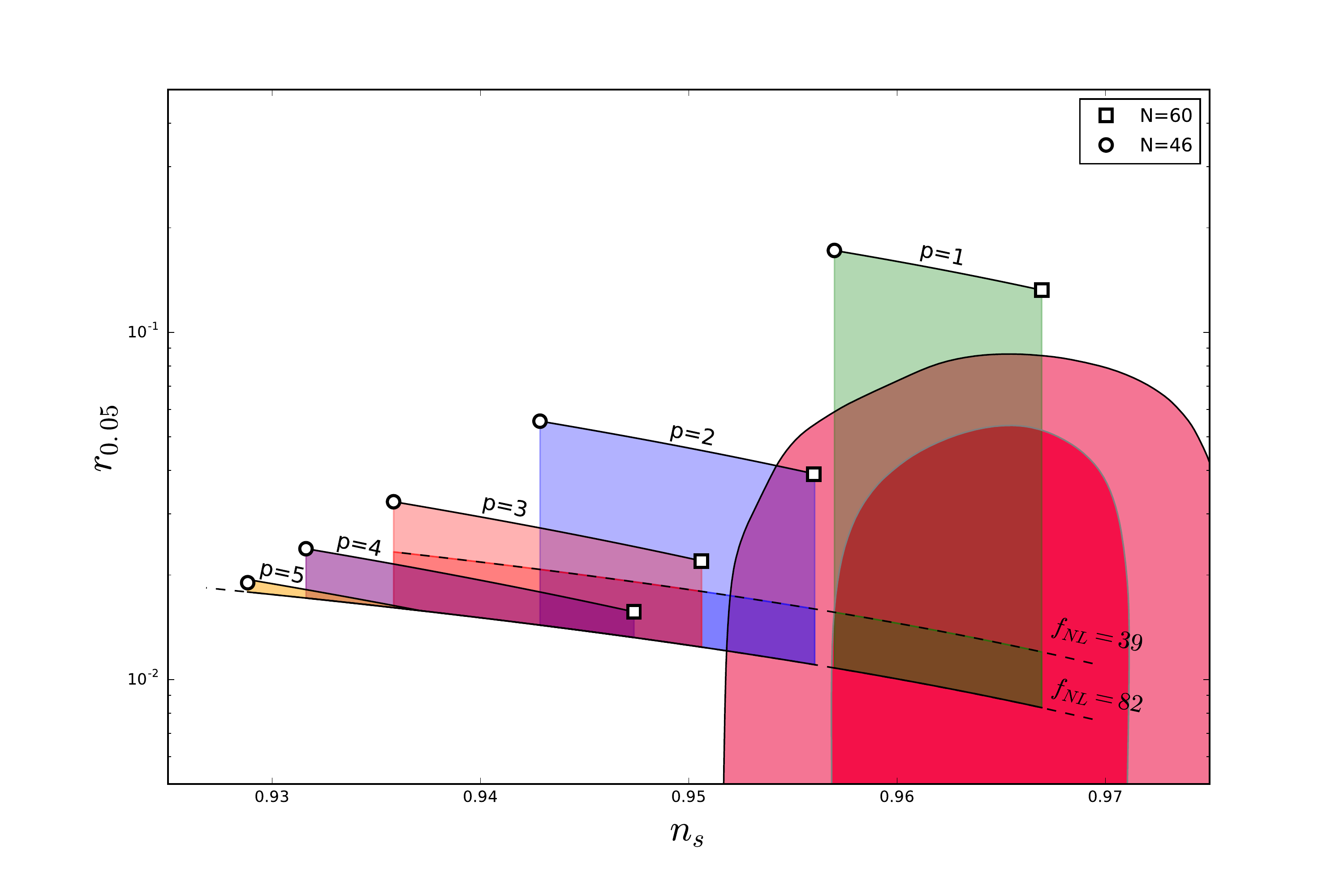}
\caption{Regions of viability for the Generalized Isokinetic scenario between the maximum and minimum $r$ values superimposed on the Planck+BICEP/KECK results, with $1\sigma$ and $2\sigma$ limits on $f^\mathrm{equil.}_\mathrm{NL}$ shown as dashed lines.}
\label{fig:isokineticconstraints}
\end{figure}

We eliminate the arbitrary mass scale $\mu$ included in $\phi_e$ by enforcing causality, $c_S < 1$, throughout the inflationary period.  Since $\gamma$ decreases monotonically with time, it will reach its minumum value at $\phi_e$, and we require this minimum to be at least 1:
\begin{equation}
\gamma (\phi_e) = 2 p^2 \left(\dfrac{M_P}{\phi_e}\right)^2\geqslant 1,
\label{eq39}
\end{equation}
so that 
\begin{equation}
\dfrac{\phi_e}{M_P}\leqslant p\sqrt{2}.
\label{eq40}
\end{equation}
This bounds the tensor/scalar ratio from above, 
\begin{equation}
r\leqslant16\epsilon^{2p/(p+1)}=16\left(1+\frac{p+1}{p}N\right)^{-2p/(p+1)}.
\label{eq41}
\end{equation} 
To put a lower bound on the tensor/scalar ratio, we must write $f^\mathrm{equil.}_\mathrm{NL}$ as a function of $r$, $p$ and $N$, rather than $c_S$. Equations (\ref{eq10}) and (\ref{eq:generalr}) establish that  
\begin{equation}
c_S=\gamma^{-1}=\frac{r}{16\epsilon},
\label{eq42}
\end{equation}
so that 
\begin{equation}
f^\mathrm{equil.}_\mathrm{NL}=\dfrac{35}{108}\left[\dfrac{256\epsilon^2}{r^2}-1\right]=\dfrac{35}{108}\left[\dfrac{256}{r^2}\left(1+\dfrac{p+1}{p}N\right)^{-2}-1\right].
\label{eq43}
\end{equation}
We invert this to find a lower bound on the tensor/scalar ratio, 
\begin{equation}
r \geq 16\left(1+\frac{p+1}{p}N\right)^{-1}\left(\dfrac{108f^\mathrm{equil.}_\mathrm{NL}}{35}+1\right)^{-1/2}.
\end{equation}
Fig. \ref{fig:isokineticconstraints} shows the constraint from Planck+BICEP/KECK: only p-values of 1 or 2 satisfy the CMB constraint on the spectral index. The non-Gaussianity constraint places a lower bound on $r$ of order $0.01$ for all cases, so any observations that $r < 0.01$ will falsify this entire class of models. 

\section{The General Slow Roll Case}
\label{sec:SlowRoll}

For comparison, we consider the general slow roll case, where the spectral index in the canonical case is determined by {\it two} slow roll parameters,
\begin{equation}
n_S = 1 - 4 \epsilon + 2 \eta.
\end{equation}
The parameters $\epsilon$ and $\eta$ are in general independent: This is in contrast to the power-law case, for which $\eta = \epsilon$, and the isokinetic case, for which $\eta = 0$. In the general case, it is possible to match the Planck constraint on $n_S$ for $\epsilon \rightarrow 0$ by taking $\eta$ negative. This is the case, for example, in ``hilltop'' models of inflation \cite{Barenboim:2016mmw}. Then the spectral index and the tensor/scalar ratio are parametrically decoupled, and the lower bound (\ref{eq:cSbound}) does not allow us to bound $r$ from below. Generalizing to non-canonical Lagrangians gives even more freedom,
\begin{equation}
n_S = 1 - 4 \epsilon + 2 \eta - 2 s,
\end{equation}
so that we are free to adjust $n_S$ to match the Planck constraint in the limit of $\epsilon \rightarrow 0$ either by $\eta < 0$ as in the case of hilltop inflation, or $s > 0$ as in the IR DBI models discussed in Sec. (\ref{sec:IRDBI}), or by a combination of the two. 

\section{Conclusions}
\label{sec:Conclusions}
In this paper, we have considered non-canonical generalizations to large-field inflation models, which generically overproduce tensors and are inconsistent with CMB constraints. Non-canonical Lagrangians are characterized by a sound speed $c_S < 1$, which results in suppression of the tensor/scalar ratio, allowing models which overproduce tensors to be brought into agreement with current constraints. We first considered non-canonical generalizations of the Power-Law inflation scenario, which is exactly solvable in both canonical and non-canonical versions. We next considered the non-canonical generalization of the simplest ``chaotic" inflation scenario with potential $V(\phi)=m^2\phi^2$, for which the field evolves with a near-constant velocity $\dot{\phi}$, called ``isokinetic'' inflation. We found that for both of these cases, the current $2\sigma$ upper bounds on  $f^\mathrm{equil.}_\mathrm{NL}$ creates a {\it lower bound} on $r$ of order $0.01$, creating a narrow window of potential viability for the models. In the power-law case, the bound only applies to ultraviolet-type DBI inflation, for which the slow roll parameter $s < 0$, and the brane evolves {\it toward} the tip of the Klebanov-Strassler throat. While such models are allowed by the Planck data alone, Planck combined with constraints from perturbative unitarity \cite{Baumann:2015nta} rule them out altogether. The original DBI model of Silverstein and Tong, for example, is of this type, with $s = - 2 \epsilon$ \cite{Silverstein:2003hf}.  Other versions of the power-law DBI inflation, however, survive, in particular the infrared-type DBI models, for which $s > 0$ \cite{Chen:2004gc,Chen:2005ad}, and the brane evolves {\it away from} the tip of the throat. In this case, there is no lower bound on the parameter $\epsilon$ and therefore no lower bound on $r$, despite the lower bound on the sound speed $c_S$. 

A tensor/scalar ratio $r \sim 0.01$ is within range of near-future Cosmic Microwave Background experiments, so that a lack of detection of tensor modes will not only rule out large-field models in their canonical form, but broad classes of non-canonical generalizations as well.  

\section*{Acknowledgments}

WHK and NKS are supported by the National Science Foundation under grant NSF-PHY-1417317. This work was performed in part at the University at Buffalo Center for Computational Research.

\bibliographystyle{JHEP}
\bibliography{Paper.bib}

\providecommand{\href}[2]{#2}\begingroup\raggedright\begin{thebibliography}{10}

\bibitem{Starobinsky:1980te}
A.~A. Starobinsky, \emph{{A New Type of Isotropic Cosmological Models Without
  Singularity}},
  \href{http://dx.doi.org/10.1016/0370-2693(80)90670-X}{\emph{Phys. Lett.} {\bf
  B91} (1980) 99--102}.

\bibitem{Sato:1981ds}
K.~Sato, \emph{{Cosmological Baryon Number Domain Structure and the First Order
  Phase Transition of a Vacuum}},
  \href{http://dx.doi.org/10.1016/0370-2693(81)90805-4}{\emph{Phys. Lett.} {\bf
  B99} (1981) 66--70}.

\bibitem{Sato:1980yn}
K.~Sato, \emph{{First Order Phase Transition of a Vacuum and Expansion of the
  Universe}}, {\emph{Mon. Not. Roy. Astron. Soc.} {\bf 195} (1981) 467--479}.

\bibitem{Kazanas:1980tx}
D.~Kazanas, \emph{{Dynamics of the Universe and Spontaneous Symmetry
  Breaking}}, \href{http://dx.doi.org/10.1086/183361}{\emph{Astrophys. J.} {\bf
  241} (1980) L59--L63}.

\bibitem{Guth:1980zm}
A.~H. Guth, \emph{{The Inflationary Universe: A Possible Solution to the
  Horizon and Flatness Problems}},
  \href{http://dx.doi.org/10.1103/PhysRevD.23.347}{\emph{Phys. Rev.} {\bf D23}
  (1981) 347--356}.

\bibitem{Linde:1981mu}
A.~D. Linde, \emph{{A New Inflationary Universe Scenario: A Possible Solution
  of the Horizon, Flatness, Homogeneity, Isotropy and Primordial Monopole
  Problems}}, \href{http://dx.doi.org/10.1016/0370-2693(82)91219-9}{\emph{Phys.
  Lett.} {\bf B108} (1982) 389--393}.

\bibitem{Albrecht:1982wi}
A.~Albrecht and P.~J. Steinhardt, \emph{{Cosmology for Grand Unified Theories
  with Radiatively Induced Symmetry Breaking}},
  \href{http://dx.doi.org/10.1103/PhysRevLett.48.1220}{\emph{Phys. Rev. Lett.}
  {\bf 48} (1982) 1220--1223}.

\bibitem{Spergel:2006hy}
{\scshape WMAP} collaboration, D.~N. Spergel et~al., \emph{{Wilkinson Microwave
  Anisotropy Probe (WMAP) three year results: implications for cosmology}},
  \href{http://dx.doi.org/10.1086/513700}{\emph{Astrophys. J. Suppl.} {\bf 170}
  (2007) 377}, [\href{http://arxiv.org/abs/astro-ph/0603449}{{\tt
  astro-ph/0603449}}].

\bibitem{Alabidi:2006qa}
L.~Alabidi and D.~H. Lyth, \emph{{Inflation models after WMAP year three}},
  \href{http://dx.doi.org/10.1088/1475-7516/2006/08/013}{\emph{JCAP} {\bf 0608}
  (2006) 013}, [\href{http://arxiv.org/abs/astro-ph/0603539}{{\tt
  astro-ph/0603539}}].

\bibitem{Seljak:2006bg}
U.~Seljak, A.~Slosar and P.~McDonald, \emph{{Cosmological parameters from
  combining the Lyman-alpha forest with CMB, galaxy clustering and SN
  constraints}},
  \href{http://dx.doi.org/10.1088/1475-7516/2006/10/014}{\emph{JCAP} {\bf 0610}
  (2006) 014}, [\href{http://arxiv.org/abs/astro-ph/0604335}{{\tt
  astro-ph/0604335}}].

\bibitem{Kinney:2006qm}
W.~H. Kinney, E.~W. Kolb, A.~Melchiorri and A.~Riotto, \emph{{Inflation model
  constraints from the Wilkinson Microwave Anisotropy Probe three-year data}},
  \href{http://dx.doi.org/10.1103/PhysRevD.74.023502}{\emph{Phys. Rev.} {\bf
  D74} (2006) 023502}, [\href{http://arxiv.org/abs/astro-ph/0605338}{{\tt
  astro-ph/0605338}}].

\bibitem{Martin:2006rs}
J.~Martin and C.~Ringeval, \emph{{Inflation after WMAP3: Confronting the
  Slow-Roll and Exact Power Spectra to CMB Data}},
  \href{http://dx.doi.org/10.1088/1475-7516/2006/08/009}{\emph{JCAP} {\bf 0608}
  (2006) 009}, [\href{http://arxiv.org/abs/astro-ph/0605367}{{\tt
  astro-ph/0605367}}].

\bibitem{Dodelson:1997hr}
S.~Dodelson, W.~H. Kinney and E.~W. Kolb, \emph{{Cosmic microwave background
  measurements can discriminate among inflation models}},
  \href{http://dx.doi.org/10.1103/PhysRevD.56.3207}{\emph{Phys. Rev.} {\bf D56}
  (1997) 3207--3215}, [\href{http://arxiv.org/abs/astro-ph/9702166}{{\tt
  astro-ph/9702166}}].

\bibitem{Kinney:1998md}
W.~H. Kinney, \emph{{Constraining inflation with cosmic microwave background
  polarization}},
  \href{http://dx.doi.org/10.1103/PhysRevD.58.123506}{\emph{Phys. Rev.} {\bf
  D58} (1998) 123506}, [\href{http://arxiv.org/abs/astro-ph/9806259}{{\tt
  astro-ph/9806259}}].

\bibitem{Ade:2015lrj}
{\scshape Planck} collaboration, P.~A.~R. Ade et~al., \emph{{Planck 2015
  results. XX. Constraints on inflation}},
  \href{http://arxiv.org/abs/1502.02114}{{\tt 1502.02114}}.

\bibitem{Ade:2015xua}
{\scshape Planck} collaboration, P.~A.~R. Ade et~al., \emph{{Planck 2015
  results. XIII. Cosmological parameters}},
  \href{http://dx.doi.org/10.1051/0004-6361/201525830}{\emph{Astron.
  Astrophys.} {\bf 594} (2016) A13},
  [\href{http://arxiv.org/abs/1502.01589}{{\tt 1502.01589}}].

\bibitem{Aghanim:2015xee}
{\scshape Planck} collaboration, N.~Aghanim et~al., \emph{{Planck 2015 results.
  XI. CMB power spectra, likelihoods, and robustness of parameters}},
  \href{http://dx.doi.org/10.1051/0004-6361/201526926}{\emph{Astron.
  Astrophys.} (2015) }, [\href{http://arxiv.org/abs/1507.02704}{{\tt
  1507.02704}}].

\bibitem{Ade:2015fwj}
{\scshape BICEP2, Keck Array} collaboration, P.~A.~R. Ade et~al., \emph{{BICEP2
  / Keck Array V: Measurements of B-mode Polarization at Degree Angular Scales
  and 150 GHz by the Keck Array}},
  \href{http://dx.doi.org/10.1088/0004-637X/811/2/126}{\emph{Astrophys. J.}
  {\bf 811} (2015) 126}, [\href{http://arxiv.org/abs/1502.00643}{{\tt
  1502.00643}}].

\bibitem{Lyth:1996im}
D.~H. Lyth, \emph{{What would we learn by detecting a gravitational wave signal
  in the cosmic microwave background anisotropy?}},
  \href{http://dx.doi.org/10.1103/PhysRevLett.78.1861}{\emph{Phys. Rev. Lett.}
  {\bf 78} (1997) 1861--1863}, [\href{http://arxiv.org/abs/hep-ph/9606387}{{\tt
  hep-ph/9606387}}].

\bibitem{Kinney:2016qyl}
W.~H. Kinney, \emph{{Limits on Entanglement Effects in the String Landscape
  from Planck and BICEP/Keck Data}},
  \href{http://arxiv.org/abs/1606.00672}{{\tt 1606.00672}}.

\bibitem{Baumann:2014cja}
D.~Baumann, D.~Green and R.~A. Porto, \emph{{B-modes and the Nature of
  Inflation}},
  \href{http://dx.doi.org/10.1088/1475-7516/2015/01/016}{\emph{JCAP} {\bf 1501}
  (2015) 016}, [\href{http://arxiv.org/abs/1407.2621}{{\tt 1407.2621}}].

\bibitem{Palma:2014faa}
G.~A. Palma and A.~Soto, \emph{{B-modes and the sound speed of primordial
  fluctuations}},
  \href{http://dx.doi.org/10.1103/PhysRevD.91.063525}{\emph{Phys. Rev.} {\bf
  D91} (2015) 063525}, [\href{http://arxiv.org/abs/1412.0033}{{\tt
  1412.0033}}].

\bibitem{Zavala:2014bda}
I.~Zavala, \emph{{Effects of the speed of sound at large-N}},
  \href{http://dx.doi.org/10.1103/PhysRevD.91.063005}{\emph{Phys. Rev.} {\bf
  D91} (2015) 063005}, [\href{http://arxiv.org/abs/1412.3732}{{\tt
  1412.3732}}].

\bibitem{Gobbetti:2015cya}
R.~Gobbetti, E.~Pajer and D.~Roest, \emph{{On the Three Primordial Numbers}},
  \href{http://dx.doi.org/10.1088/1475-7516/2015/09/058}{\emph{JCAP} {\bf 1509}
  (2015) 058}, [\href{http://arxiv.org/abs/1505.00968}{{\tt 1505.00968}}].

\bibitem{Kohri:2007gq}
K.~Kohri, C.-M. Lin and D.~H. Lyth, \emph{{More hilltop inflation models}},
  \href{http://dx.doi.org/10.1088/1475-7516/2007/12/004}{\emph{JCAP} {\bf 0712}
  (2007) 004}, [\href{http://arxiv.org/abs/0707.3826}{{\tt 0707.3826}}].

\bibitem{Martin:2013tda}
J.~Martin, C.~Ringeval and V.~Vennin, \emph{{Encyclopædia Inflationaris}},
  \href{http://dx.doi.org/10.1016/j.dark.2014.01.003}{\emph{Phys. Dark Univ.}
  {\bf 5-6} (2014) 75--235}, [\href{http://arxiv.org/abs/1303.3787}{{\tt
  1303.3787}}].

\bibitem{Barenboim:2013wra}
G.~Barenboim, E.~J. Chun and H.~M. Lee, \emph{{Coleman-Weinberg Inflation in
  light of Planck}},
  \href{http://dx.doi.org/10.1016/j.physletb.2014.01.039}{\emph{Phys. Lett.}
  {\bf B730} (2014) 81--88}, [\href{http://arxiv.org/abs/1309.1695}{{\tt
  1309.1695}}].

\bibitem{Coone:2015fha}
D.~Coone, D.~Roest and V.~Vennin, \emph{{The Hubble Flow of Plateau
  Inflation}},
  \href{http://dx.doi.org/10.1088/1475-7516/2015/11/010}{\emph{JCAP} {\bf 1511}
  (2015) 010}, [\href{http://arxiv.org/abs/1507.00096}{{\tt 1507.00096}}].

\bibitem{Huang:2015cke}
Q.-G. Huang, K.~Wang and S.~Wang, \emph{{Inflation model constraints from data
  released in 2015}},
  \href{http://dx.doi.org/10.1103/PhysRevD.93.103516}{\emph{Phys. Rev.} {\bf
  D93} (2016) 103516}, [\href{http://arxiv.org/abs/1512.07769}{{\tt
  1512.07769}}].

\bibitem{Vennin:2015egh}
V.~Vennin, K.~Koyama and D.~Wands, \emph{{Inflation with an extra light scalar
  field after Planck}},
  \href{http://dx.doi.org/10.1088/1475-7516/2016/03/024}{\emph{JCAP} {\bf 1603}
  (2016) 024}, [\href{http://arxiv.org/abs/1512.03403}{{\tt 1512.03403}}].

\bibitem{Barenboim:2016mmw}
G.~Barenboim, W.-I. Park and W.~H. Kinney, \emph{{Eternal Hilltop Inflation}},
  \href{http://dx.doi.org/10.1088/1475-7516/2016/05/030}{\emph{JCAP} {\bf 1605}
  (2016) 030}, [\href{http://arxiv.org/abs/1601.08140}{{\tt 1601.08140}}].

\bibitem{Kehagias:2013mya}
A.~Kehagias, A.~M. Dizgah and A.~Riotto, \emph{{Remarks on the Starobinsky
  model of inflation and its descendants}},
  \href{http://dx.doi.org/10.1103/PhysRevD.89.043527}{\emph{Phys. Rev.} {\bf
  D89} (2014) 043527}, [\href{http://arxiv.org/abs/1312.1155}{{\tt
  1312.1155}}].

\bibitem{Bastero-Gil:2016qru}
M.~Bastero-Gil, A.~Berera, R.~O. Ramos and J.~G. Rosa, \emph{{Warm Little
  Inflaton}},
  \href{http://dx.doi.org/10.1103/PhysRevLett.117.151301}{\emph{Phys. Rev.
  Lett.} {\bf 117} (2016) 151301}, [\href{http://arxiv.org/abs/1604.08838}{{\tt
  1604.08838}}].

\bibitem{Kinney:2007ag}
W.~H. Kinney and K.~Tzirakis, \emph{{Quantum modes in DBI inflation: exact
  solutions and constraints from vacuum selection}},
  \href{http://dx.doi.org/10.1103/PhysRevD.77.103517}{\emph{Phys. Rev.} {\bf
  D77} (2008) 103517}, [\href{http://arxiv.org/abs/0712.2043}{{\tt
  0712.2043}}].

\bibitem{Tzirakis:2008qy}
K.~Tzirakis and W.~H. Kinney, \emph{{Non-canonical generalizations of slow-roll
  inflation models}},
  \href{http://dx.doi.org/10.1088/1475-7516/2009/01/028}{\emph{JCAP} {\bf 0901}
  (2009) 028}, [\href{http://arxiv.org/abs/0810.0270}{{\tt 0810.0270}}].

\bibitem{Ade:2015ava}
{\scshape Planck} collaboration, P.~A.~R. Ade et~al., \emph{{Planck 2015
  results. XVII. Constraints on primordial non-Gaussianity}},
  \href{http://arxiv.org/abs/1502.01592}{{\tt 1502.01592}}.

\bibitem{Kinney:2002qn}
W.~H. Kinney, \emph{{Inflation: Flow, fixed points and observables to arbitrary
  order in slow roll}},
  \href{http://dx.doi.org/10.1103/PhysRevD.66.083508}{\emph{Phys. Rev.} {\bf
  D66} (2002) 083508}, [\href{http://arxiv.org/abs/astro-ph/0206032}{{\tt
  astro-ph/0206032}}].

\bibitem{Peiris:2007gz}
H.~V. Peiris, D.~Baumann, B.~Friedman and A.~Cooray, \emph{{Phenomenology of
  D-Brane Inflation with General Speed of Sound}},
  \href{http://dx.doi.org/10.1103/PhysRevD.76.103517}{\emph{Phys. Rev.} {\bf
  D76} (2007) 103517}, [\href{http://arxiv.org/abs/0706.1240}{{\tt
  0706.1240}}].

\bibitem{Liddle:1994dx}
A.~R. Liddle, P.~Parsons and J.~D. Barrow, \emph{{Formalizing the slow roll
  approximation in inflation}},
  \href{http://dx.doi.org/10.1103/PhysRevD.50.7222}{\emph{Phys. Rev.} {\bf D50}
  (1994) 7222--7232}, [\href{http://arxiv.org/abs/astro-ph/9408015}{{\tt
  astro-ph/9408015}}].

\bibitem{Kinney:2009vz}
W.~H. Kinney, \emph{{TASI Lectures on Inflation}},
  \href{http://arxiv.org/abs/0902.1529}{{\tt 0902.1529}}.

\bibitem{Silverstein:2003hf}
E.~Silverstein and D.~Tong, \emph{{Scalar speed limits and cosmology:
  Acceleration from D-cceleration}},
  \href{http://dx.doi.org/10.1103/PhysRevD.70.103505}{\emph{Phys. Rev.} {\bf
  D70} (2004) 103505}, [\href{http://arxiv.org/abs/hep-th/0310221}{{\tt
  hep-th/0310221}}].

\bibitem{Alishahiha:2004eh}
M.~Alishahiha, E.~Silverstein and D.~Tong, \emph{{DBI in the sky}},
  \href{http://dx.doi.org/10.1103/PhysRevD.70.123505}{\emph{Phys. Rev.} {\bf
  D70} (2004) 123505}, [\href{http://arxiv.org/abs/hep-th/0404084}{{\tt
  hep-th/0404084}}].

\bibitem{Chen:2006nt}
X.~Chen, M.-x. Huang, S.~Kachru and G.~Shiu, \emph{{Observational signatures
  and non-Gaussianities of general single field inflation}},
  \href{http://dx.doi.org/10.1088/1475-7516/2007/01/002}{\emph{JCAP} {\bf 0701}
  (2007) 002}, [\href{http://arxiv.org/abs/hep-th/0605045}{{\tt
  hep-th/0605045}}].

\bibitem{Spalinski:2007qy}
M.~Spalinski, \emph{{A Consistency Relation for Power Law Inflation in DBI
  models}}, \href{http://dx.doi.org/10.1016/j.physletb.2007.05.041}{\emph{Phys.
  Lett.} {\bf B650} (2007) 313--316},
  [\href{http://arxiv.org/abs/hep-th/0703248}{{\tt hep-th/0703248}}].

\bibitem{ArmendarizPicon:1999rj}
C.~Armendariz-Picon, T.~Damour and V.~F. Mukhanov, \emph{{k - inflation}},
  \href{http://dx.doi.org/10.1016/S0370-2693(99)00603-6}{\emph{Phys. Lett.}
  {\bf B458} (1999) 209--218}, [\href{http://arxiv.org/abs/hep-th/9904075}{{\tt
  hep-th/9904075}}].

\bibitem{Cheung:2007st}
C.~Cheung, P.~Creminelli, A.~L. Fitzpatrick, J.~Kaplan and L.~Senatore,
  \emph{{The Effective Field Theory of Inflation}},
  \href{http://dx.doi.org/10.1088/1126-6708/2008/03/014}{\emph{JHEP} {\bf 03}
  (2008) 014}, [\href{http://arxiv.org/abs/0709.0293}{{\tt 0709.0293}}].

\bibitem{Baumann:2015nta}
D.~Baumann, D.~Green, H.~Lee and R.~A. Porto, \emph{{Signs of Analyticity in
  Single-Field Inflation}},
  \href{http://dx.doi.org/10.1103/PhysRevD.93.023523}{\emph{Phys. Rev.} {\bf
  D93} (2016) 023523}, [\href{http://arxiv.org/abs/1502.07304}{{\tt
  1502.07304}}].

\bibitem{Baumann:2011su}
D.~Baumann and D.~Green, \emph{{Equilateral Non-Gaussianity and New Physics on
  the Horizon}},
  \href{http://dx.doi.org/10.1088/1475-7516/2011/09/014}{\emph{JCAP} {\bf 1109}
  (2011) 014}, [\href{http://arxiv.org/abs/1102.5343}{{\tt 1102.5343}}].

\bibitem{Gwyn:2012mw}
R.~Gwyn, G.~A. Palma, M.~Sakellariadou and S.~Sypsas, \emph{{Effective field
  theory of weakly coupled inflationary models}},
  \href{http://dx.doi.org/10.1088/1475-7516/2013/04/004}{\emph{JCAP} {\bf 1304}
  (2013) 004}, [\href{http://arxiv.org/abs/1210.3020}{{\tt 1210.3020}}].

\bibitem{Chimento:2007es}
L.~P. Chimento and R.~Lazkoz, \emph{{Bridging geometries and potentials in DBI
  cosmologies}}, \href{http://dx.doi.org/10.1007/s10714-008-0637-1}{\emph{Gen.
  Rel. Grav.} {\bf 40} (2008) 2543--2555},
  [\href{http://arxiv.org/abs/0711.0712}{{\tt 0711.0712}}].

\bibitem{Spalinski:2007kt}
M.~Spalinski, \emph{{On the Slow Roll Expansion for Brane Inflation}},
  \href{http://dx.doi.org/10.1088/1475-7516/2007/04/018}{\emph{JCAP} {\bf 0704}
  (2007) 018}, [\href{http://arxiv.org/abs/hep-th/0702118}{{\tt
  hep-th/0702118}}].

\bibitem{Bessada:2009ns}
D.~Bessada, W.~H. Kinney, D.~Stojkovic and J.~Wang, \emph{{Tachyacoustic
  Cosmology: An Alternative to Inflation}},
  \href{http://dx.doi.org/10.1103/PhysRevD.81.043510}{\emph{Phys. Rev.} {\bf
  D81} (2010) 043510}, [\href{http://arxiv.org/abs/0908.3898}{{\tt
  0908.3898}}].

\bibitem{Chen:2004gc}
X.~Chen, \emph{{Multi-throat brane inflation}},
  \href{http://dx.doi.org/10.1103/PhysRevD.71.063506}{\emph{Phys. Rev.} {\bf
  D71} (2005) 063506}, [\href{http://arxiv.org/abs/hep-th/0408084}{{\tt
  hep-th/0408084}}].

\bibitem{Chen:2005ad}
X.~Chen, \emph{{Inflation from warped space}},
  \href{http://dx.doi.org/10.1088/1126-6708/2005/08/045}{\emph{JHEP} {\bf 08}
  (2005) 045}, [\href{http://arxiv.org/abs/hep-th/0501184}{{\tt
  hep-th/0501184}}].

\end{thebibliography}\endgroup

\end{document}